%% file: main.tex
\documentclass[10pt]{article}
\usepackage[preprint]{rlc}
\usepackage{url}
\usepackage{xurl}
\usepackage{graphicx}
\usepackage{amsmath}
\usepackage{booktabs}

\usepackage{fancyhdr}
\usepackage{tikz}
\usepackage{enumitem}
\usetikzlibrary{positioning, arrows.meta, fit, backgrounds, calc}

\title{\centering AWCP: A Workspace Delegation Protocol for Deep-Engagement Collaboration across Remote Agents}

\author{
\parbox{\textwidth}{
\centering
Xiaohang Nie$^{1,2,\dagger}$, Zihan Guo$^{2,3,\dagger}$, Youliang Chen$^{4}$, Yuanjian Zhou$^{2,*}$, Weinan Zhang$^{2,5,*}$
}
\\
$^{1}$ Harbin Institute of Technology\quad$^{2}$ Shanghai Innovation Institute\quad$^{3}$ Sun Yat-sen University
\\
$^{4}$ Tongji University\quad$^{5}$ Shanghai Jiao Tong University
\\
$^{\dagger}$ Equal contribution.\quad$^{*}$~Corresponding author.
\\
\texttt{253108020043@sii.edu.cn}\quad\texttt{wnzhang@sjtu.edu.cn}
}

\begin{document}

\maketitle

\begin{abstract}
The rapid evolution of Large Language Model~(LLM)-based autonomous agents is reshaping the digital landscape toward an emerging Agentic Web, where increasingly specialized agents must collaborate to accomplish complex tasks.
However, existing collaboration paradigms are constrained to message passing, leaving execution environments as isolated silos.
This creates a \textit{context gap}: agents cannot directly manipulate files or invoke tools in a peer's environment, and must instead resort to costly, error-prone environment reconstruction.
We introduce the Agent Workspace Collaboration Protocol~(AWCP), which bridges this gap through temporary workspace delegation inspired by the Unix philosophy that \textit{everything is a file}.
AWCP decouples a lightweight control plane from pluggable transport mechanisms, allowing a Delegator to project its workspace to a remote Executor, who then operates on the shared files directly with unmodified local toolchains.
We provide a fully open-source reference implementation with MCP tool integration and validate the protocol through live demonstrations of asymmetric collaboration, where agents with complementary capabilities cooperate through delegated workspaces.
By establishing the missing workspace layer in the agentic protocol stack, AWCP paves the way for a universally interoperable agent ecosystem in which collaboration transcends message boundaries.
The protocol and reference implementation are publicly available at \url{https://github.com/SII-Holos/awcp}.
\end{abstract}

\section{Introduction}
\label{sec:introduction}

\begin{figure}[t]
\centering
\includegraphics[width=0.85\textwidth]{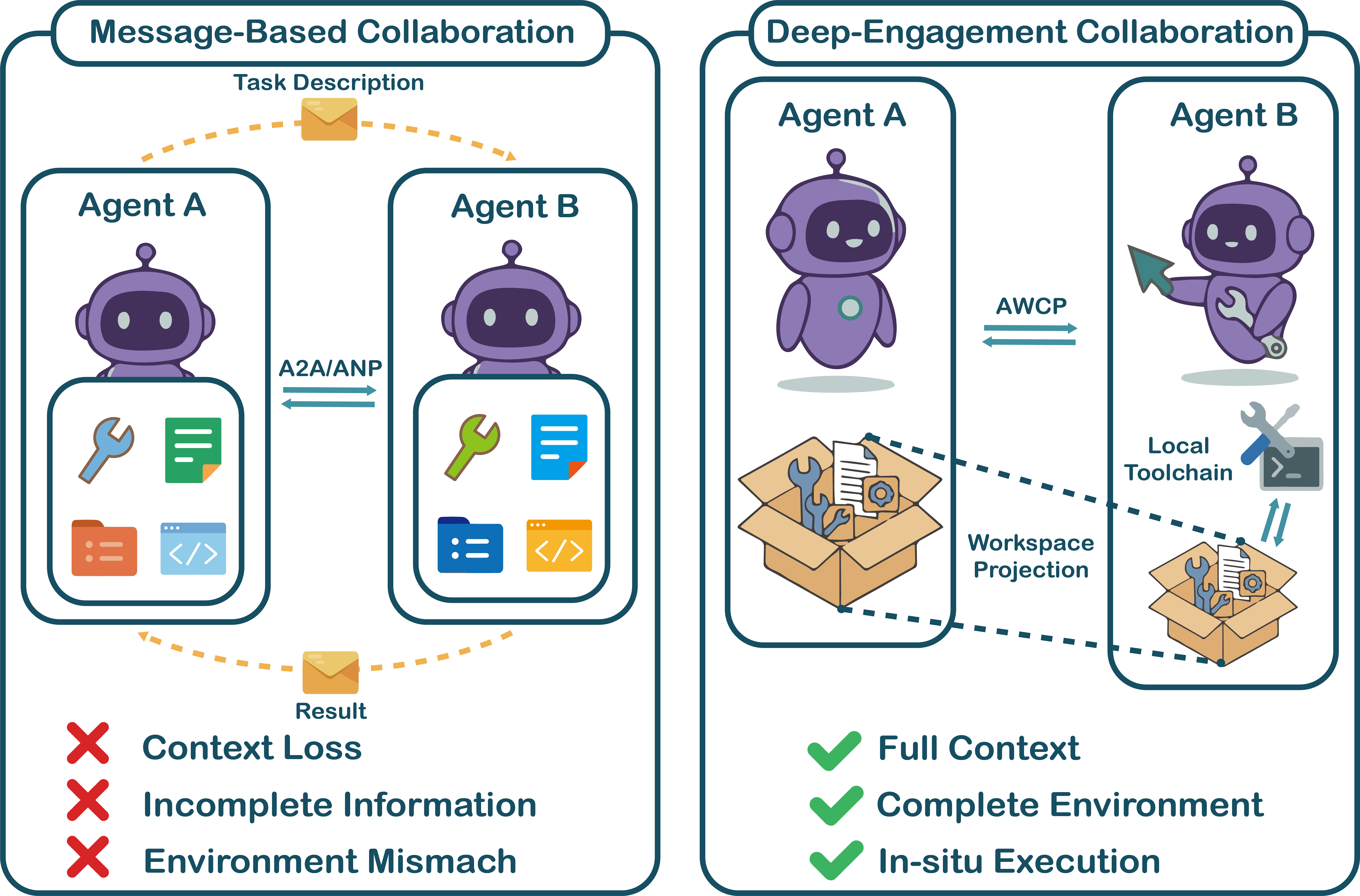}
\caption{Two paradigms for inter-agent collaboration. (a)~In message-based protocols such as A2A and ANP, agents exchange serialized task descriptions and results; the receiving agent operates on detached artifacts without access to the originator's execution environment, leading to well-documented failure modes including context loss, incomplete information transfer, and environment mismatch~\protect\citep{cemri2025mast}. (b)~In AWCP's deep-engagement model, the Delegator projects its workspace to the Executor, who operates directly on the original files using its local toolchain, preserving full context, providing the complete environment, and enabling in-situ execution.}
\label{fig:paradigm-comparison}
\end{figure}

As autonomous agents evolve from conversational chatbots to pervasive system assistants, each accumulates domain-specific expertise that remains confined within its siloed execution environment~\citep{wang2024survey, yang2025agenticweb}.
This tension between deepening vertical expertise and persistent horizontal disconnection demands a redefinition of agent collaboration, which entails shifting from superficial message passing to \textit{deep-engagement} interoperability at the system level~\citep{ehtesham2025survey}, as illustrated in Figure~\ref{fig:paradigm-comparison}.

Recent agentic protocols have begun to address distinct facets of this interaction landscape.
The Model Context Protocol~(MCP) standardizes the \textit{agent--tool} boundary, connecting agents to external tools, data sources, and APIs through structured function calls~\citep{anthropic2024mcp}.
The Agent-to-Agent Protocol~(A2A) tackles the \textit{agent--agent} boundary, enabling agents to discover peers through Agent Cards and coordinate tasks via structured message exchange~\citep{google2025a2a}.
Collectively, these protocols cover tool access and task-level coordination.
However, both operate strictly at the message layer, e.g., MCP returns discrete function outputs, and A2A exchanges structured payloads.
Neither empowers an agent to perform native operations like \texttt{ls}, \texttt{cat}, or \texttt{git commit} directly within a peer's workspace.
As illustrated in Figure~\ref{fig:protocol-landscape}, a critical gap persists at the \textit{agent--workspace} boundary, the layer where agents require direct, filesystem-level access to each other's execution environments.

\input{figs/protocol-landscape}

To make this gap concrete, consider a coding agent that has completed a feature branch and needs a remote security-auditing agent to review it.
Via MCP, the auditor might lint files individually, but it lacks the systemic context required to navigate the repository, resolve cross-file dependencies, or execute project-wide static analysis.
Under A2A, it could receive selected sources as task attachments, which arrive as flat snapshots stripped of the build system, version history, and test infrastructure.
What the agent actually needs is to \textit{project its workspace} to the auditor, granting temporary, filesystem-level access so that the remote agent can operate with its native toolchains as if working locally.
Indeed, recent multi-agent systems have begun to approximate this pattern through ad hoc mechanisms such as per-agent sandboxes and shared Git repositories, yet even web-scale collaboration platforms such as Holos~\citep{holos2026} that orchestrate millions of heterogeneous agents lack a standardized protocol for cross-agent workspace delegation.

We introduce the Agent Workspace Collaboration Protocol~(AWCP), formalizing temporary workspace delegation via a standardized framework that enables \textit{deep-engagement collaboration} between autonomous agents.
Inspired by the Unix philosophy that \textit{everything is a file}~\citep{ritchie1974unix}, AWCP establishes a \textit{files-as-interface} paradigm for Delegator--Executor collaboration.
This design reflects a fundamental observation: coding agents read files, write patches, and run tests~\citep{jimenez2024swebench, yang2024sweagent}; vision agents inspect image directories; compliance agents stamp PDF documents.
The filesystem is not merely one possible interface for collaboration, but it is the medium through which agents already interact with their computational surroundings.
By abstracting collaboration into file operations, AWCP enables a Delegator to project its environment to a remote Executor, who then works on the shared files with unmodified local toolchains.
The protocol decouples a lightweight control plane from pluggable transport adapters, and integrates naturally into the existing protocol ecosystem: Delegator agents access AWCP through MCP tool interfaces or skill modules, while Executor discovery and signaling can leverage A2A, the Agent Network Protocol~(ANP), or direct HTTP endpoints~(Figure~\ref{fig:protocol-landscape}).

Our contributions are as follows:
\begin{enumerate}[nosep,leftmargin=*]
    \item \textbf{A new paradigm for agent collaboration.} We propose AWCP, the first protocol to formalize workspace delegation, shifting the focus from message-centric interaction to environment-level \textit{deep-engagement} collaboration via a \textit{files-as-interface} abstraction.
    \item \textbf{A universal and decoupled architecture.} We design a modular framework that separates the control logic from transport mechanisms, providing a standardized backbone for deep-engagement across heterogeneous agents and diverse network environments.
    \item \textbf{Foundational infrastructure for the Agentic Web.} We provide an open-source implementation and showcase asymmetric collaboration, demonstrating the viability of a universally interoperable agentic ecosystem.
\end{enumerate}

The remainder of this paper is organized as follows: 
Section 2 provides a review of the related research background. 
Section 3 introduces the AWCP framework and its delegation model, followed by a detailed discussion of the protocol specification and engineering implementation in Section 4. 
Section 5 then demonstrates the framework's effectiveness through use cases in cross-modal dataset curation and compliance stamping. Finally, Section 6 summarizes our findings and discusses future directions.

\section{Related Work}
\label{sec:related}

AWCP builds upon and extends three research threads: agent communication protocols that define how agents exchange information, multi-agent frameworks that orchestrate collaborative workflows, and distributed file systems that enable transparent remote resource access.

\subsection{Agent Communication Protocols}

Enabling effective communication among autonomous agents has been a long-standing research challenge.
The Foundation for Intelligent Physical Agents~(FIPA) standardized an Agent Communication Language based on speech act theory, defining 22 communicative acts~(inform, request, propose, etc.) with formal semantics~\citep{fipa2002acl}.
The Contract Net Protocol formalized task delegation through announcement, bidding, and awarding phases~\citep{smith1980contract}.
While foundational, these classical protocols were designed for rule-based agents and rely on rigid logical content languages that are less compatible with the stochastic nature of modern LLMs.

The emergence of LLM-based agents has catalyzed a new generation of protocols~\citep{yang2025survey, kong2025survey}.
MCP establishes a client-server architecture connecting AI applications to external tools and data sources through JSON-RPC, exposing structured primitives for tool invocation, resource access, and prompt templates~\citep{anthropic2024mcp, ehtesham2025survey}.
A2A addresses peer-to-peer agent coordination, introducing Agent Cards for capability discovery and a task lifecycle model~(submitted, working, completed) for managing delegated work through structured message exchange~\citep{google2025a2a}.
ANP operates at the network layer, proposing a three-tier architecture for agent identity via W3C Decentralized Identifiers, meta-protocol negotiation, and capability discovery~\citep{chang2025anp}.

Despite this rapid progress, all existing protocols share a fundamental limitation: they operate at the \textit{message layer}.
MCP transmits structured function calls and returns discrete outputs; A2A exchanges task messages and file attachments as payloads; ANP negotiates communication channels.
AWCP addresses the orthogonal \textit{workspace layer}, enabling agents to interact through projected filesystem environments rather than serialized payloads.

\subsection{Multi-Agent Frameworks}

A parallel line of research develops frameworks for orchestrating multiple LLM-based agents.
Early systems established the foundational patterns.
AutoGen~\citep{wu2024autogen} and CAMEL~\citep{li2023camel} explore flexible multi-agent dialogue, the former through conversable agents with multi-turn exchange and the latter through inception-prompted role-playing.
MetaGPT~\citep{hong2024metagpt} takes a more structured approach, encoding Standardized Operating Procedures into role-specialized assembly-line pipelines.
Subsequent systems have scaled these ideas further: ChatDev~\citep{qian2024chatdev} organizes agents into a virtual software company, while AgentScope~\citep{gao2025agentscope} evolves into a developer-centric platform with MCP integration and runtime sandboxing.

Since 2025, the field has moved beyond these centralized, role-based designs.
One direction pursues decentralization: Symphony~\citep{wang2025symphony} eliminates the central orchestrator altogether, coordinating lightweight LLMs across consumer-grade hardware through a capability ledger and Beacon-selection protocol.
Another direction pursues structural adaptivity: AdaptOrch~\citep{yu2026adaptorch} formalizes topology selection among parallel, sequential, hierarchical, and hybrid orchestration patterns, showing that how agents are composed now matters more than which model each agent runs.
A third direction automates the design of agent systems themselves, as exemplified by InfiAgent~\citep{yu2025infiagent}, whose ``agent-as-a-tool'' mechanism lets a self-evolving DAG of agents restructure itself in response to new tasks.

Most relevant to AWCP is a nascent trend toward workspace-level coordination in multi-agent software engineering.
Several systems have converged on version control as the coordination substrate: EvoGit~\citep{huang2025evogit} lets a population of coding agents coordinate exclusively through a shared Git repository, treating the version graph as an implicit communication channel, while AgentGit~\citep{li2025agentgit} applies commit-revert-branch primitives to manage workflow state across agents.
Others explore alternative shared-state mechanisms: Agyn~\citep{benkovich2026agyn} provisions isolated sandboxes per agent and merges results through pull requests, and CodeCRDT~\citep{pugachev2025codecrdt} uses Conflict-Free Replicated Data Types for lock-free concurrent code generation.
At the protocol level, SEMAP~\citep{mao2025semap} proposes lifecycle-guided execution with behavioral contracts atop A2A, reducing coordination failures by up to 69.6\%.

While these systems demonstrate that workspace-level coordination is a natural and effective paradigm, each implements its own ad hoc mechanism (e.g., Git graphs, sandbox-PR loops, CRDTs, or message contracts) without a unified protocol for workspace lifecycle governance.
AWCP addresses this gap: rather than competing with existing frameworks, it provides a standardized workspace delegation primitive with explicit lifecycle management (negotiation, provisioning, execution, completion) and transport-agnostic file synchronization that any orchestration system can adopt.

\subsection{Distributed File Systems and the Files-as-Interface Paradigm}

AWCP's \textit{files-as-interface} design draws directly from the Unix philosophy that \textit{everything is a file}~\citep{ritchie1974unix}.
Plan~9 from Bell Labs extended this principle to distributed systems, introducing per-process namespaces and the 9P protocol to make all resources, including network interfaces, graphics, remote processes, accessible as file hierarchies~\citep{pike1995plan9}.
The Sun Network Filesystem~(NFS) demonstrated that applications can operate on remote files transparently, establishing the stateless RPC-based model for network file access~\citep{sandberg1985nfs}.
FUSE~(Filesystem in Userspace) further democratized custom filesystem development, enabling user-space implementations such as SSHFS with moderate overhead suitable for many workloads~\citep{vangoor2017fuse}.

This four-decade lineage establishes that filesystem abstractions can serve as universal interfaces for cross-boundary resource access.
Modern AI agents reinforce this observation: the tasks on which they are most commonly evaluated, from patch generation~\citep{jimenez2024swebench} and repository-level code editing~\citep{yang2024sweagent} to document processing, ultimately reduce to reading, writing, and organizing files.
The filesystem is not merely one possible interface for agent collaboration; it is the native medium through which agents already interact with their computational surroundings.
AWCP's transport layer builds on these proven foundations, adapting the ``filesystem as universal interface'' paradigm to cross-agent delegation across diverse network conditions.

Among contemporary agent systems, OpenHands~\citep{wang2025openhands} and SWE-Agent~\citep{yang2024sweagent} address the closely related problem of granting AI agents access to execution environments, through container-based sandboxing and specialized Agent-Computer Interfaces, respectively.
EnvX~\citep{chen2025envx} extends this direction by transforming GitHub repositories into interactive agents through a three-phase environment initialization lifecycle (dependency resolution, data provisioning, validation) that closely parallels AWCP's provisioning flow.
At the conceptual level, Hassan et al.~\citep{hassan2025sase} articulate a vision for Structured Agentic Software Engineering in which an \textit{Agent Execution Environment} provides a formal digital workspace for agent teams, though they leave the protocol-level realization as future work.
These approaches share AWCP's recognition that agents need filesystem-level access, but they focus on provisioning single-agent environments or articulating theoretical frameworks rather than enabling cross-agent workspace delegation with a concrete protocol.
AWCP fills precisely this role, contributing the protocol layer that lets one agent project its workspace to another, independent of the execution environment either agent uses internally.

\section{AWCP Framework}
\label{sec:framework}

AWCP's core mechanism is workspace projection: a Delegator exposes a selected directory tree to a remote Executor, who works on it with local tools, and the protocol synchronizes results upon completion.
This section presents the framework design: a modular architecture organizing cross-node responsibilities~(\S\ref{subsec:architecture}) and dual state machines formalizing the delegation lifecycle~(\S\ref{subsec:lifecycle}).

\subsection{Architecture}
\label{subsec:architecture}

\begin{figure}[t]
    \centering
    \includegraphics[width=0.85\textwidth, trim=1cm 1cm 1cm 1cm, clip]{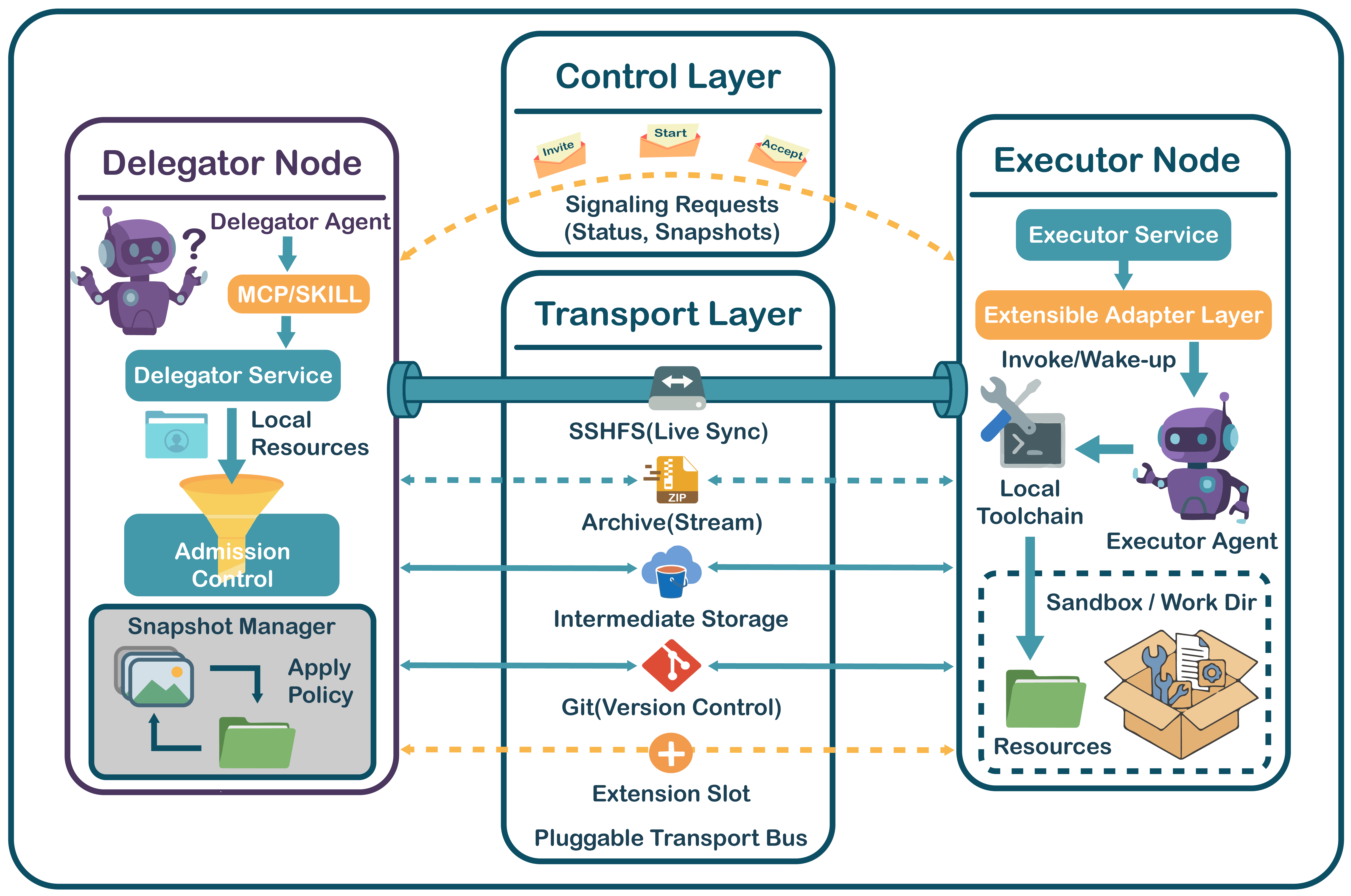}
    \caption{AWCP architecture. The control layer handles signaling and state synchronization between the Delegator and Executor, while
    the transport layer provides pluggable adapters (SSHFS, Archive, Storage, Git) for data exchange. Both nodes run dedicated AWCP services while agents operate with their native toolchains.}
    \label{fig:architecture}
\end{figure}

As illustrated in Figure~\ref{fig:architecture}, AWCP employs a dual-layer architecture to facilitate interaction between a Delegator and an Executor, both being autonomous agents.
Each node runs a dedicated AWCP service responsible for protocol signaling and resource orchestration.

On the \textbf{Delegator} side, agents interface with the \textit{Delegator Service} through MCP~\citep{anthropic2024mcp} or skill integrations~\citep{anthropic2025agentskills}.
An \textit{Admission Control} module enforces configurable policies to ensure that only well-scoped workspaces are exposed, while a \textit{Snapshot Manager} reconciles results upon task completion through automated application or manual review.

On the \textbf{Executor} side, the \textit{Executor Service} receives incoming delegation requests and provisions session-specific directories.
An \textit{Extensible Adapter Layer} decouples the service from specific agent implementations, facilitating integration with diverse backends such as A2A-compliant runtimes~\citep{google2025a2a}.
Once the workspace is mounted or extracted, the Executor agent applies native toolchains directly to delegated files, achieving the \textit{on-site access} that underpins AWCP's design.

The Delegator and Executor services communicate through two protocol layers whose strict separation is central to the design.

The \textbf{Control Layer} manages the delegation lifecycle via lightweight HTTP requests and Server-Sent Events~(SSE)~\citep{whatwg2026sse}, handling session negotiation, lease enforcement, and status synchronization.
The \textbf{Transport Layer} provisions the data plane through pluggable adapters, each consisting of a complementary pair: one component on the Delegator side (packaging workspaces and applying results) and one on the Executor side (provisioning access and capturing snapshots).
This paired design reflects the inherently asymmetric roles in workspace delegation.
Each adapter further declares its \textit{capabilities}, namely whether it supports live synchronization or snapshot-based exchange, enabling the control layer to adapt its reconciliation strategy at runtime.

This decoupling ensures that the control logic remains invariant regardless of whether files are projected via a live FUSE mount, shipped as a ZIP archive, exchanged through cloud storage, or synchronized via Git.
Network topology, workspace scale, and deployment constraints thus influence only adapter selection rather than the coordination protocol itself.

\subsection{Delegation Lifecycle}
\label{subsec:lifecycle}

\input{figs/awcp-sequence-diagram}

AWCP formalizes the delegation lifecycle through two cooperating state machines, one located on each side of the protocol, which are synchronized by a four-phase message exchange.
Figure~\ref{fig:awcp-protocol} illustrates the corresponding message sequence.

\paragraph{Dual state machines.}
The protocol separates delegation governance into two distinct state machines that operate independently yet synchronize at well-defined points through protocol messages.

The \textit{DelegationStateMachine} operates on the Delegator side and tracks nine states: an initial state \texttt{created}, four intermediate states (\texttt{invited}, \texttt{accepted}, \texttt{started}, \texttt{running}), which represent progression through the negotiation, provisioning, and execution phases, and four terminal states (\texttt{completed}, \texttt{error}, \texttt{cancelled}, \texttt{expired}).
Transitions are triggered by nine event types, each governed by transition guards that restrict activation to specific source states.
The happy path follows the sequence $\texttt{created} \to \texttt{invited} \to \texttt{accepted} \to \texttt{started} \to \texttt{running} \to \texttt{completed}$.

From any non-terminal state, an \texttt{ERROR} or \texttt{CANCEL} event transitions the machine to the corresponding terminal state; an \texttt{EXPIRE} event is additionally valid from \texttt{invited}, \texttt{accepted}, or \texttt{running} when the lease duration elapses.

The \textit{AssignmentStateMachine} operates on the Executor side with a deliberately minimal four-state design: \texttt{pending} (awaiting the \texttt{START} message after accepting an invitation), \texttt{active} (task executing), \texttt{completed}, and \texttt{error}.
This asymmetry reflects the protocol's division of responsibility: the Delegator manages the full lifecycle including lease enforcement and snapshot reconciliation, while the Executor focuses exclusively on task execution within the provisioned workspace.
Figure~\ref{fig:state-machines} presents the formal state transition structure of both machines.

\input{figs/state-machines}

Three protocol messages serve as synchronization points between the two machines.
When the Executor sends \texttt{ACCEPT}, the Delegator transitions from \texttt{invited} to \texttt{accepted} while the Executor creates a new assignment in the \texttt{pending} state.
When the Delegator sends \texttt{START}, it transitions to \texttt{started} and the Executor advances from \texttt{pending} to \texttt{active}.
When the Executor emits \texttt{DONE}, it reaches \texttt{completed} and the Delegator transitions from \texttt{running} to \texttt{completed}.
This synchronization design ensures that both parties maintain consistent views of the delegation progress without requiring a shared state store.

\paragraph{Four-phase protocol flow.}
The lifecycle unfolds across four sequential phases.

During \textit{Negotiation}, the Delegator proposes a task via an \texttt{INVITE} message.
The Executor evaluates the proposal against local policies and responds with \texttt{ACCEPT}, optionally narrowing the terms (e.g., capping the lease duration), or \texttt{ERROR} to decline.

\textit{Provisioning} activates the data plane.
The Delegator packages transport-specific credentials into a \texttt{START} message, and the Executor provisions workspace access upon receipt.

\textit{Execution} is the functional core of the protocol.
The Executor operates on the delegated workspace using its native toolchains, streaming progress and intermediate results back to the Delegator via SSE.
Upon completion, the Executor tears down transport resources and signals \texttt{DONE}.

Finally, \textit{Completion} reconciles results: the Delegator processes any received snapshots, acknowledges receipt, and both parties perform a two-phase cleanup, which involves gracefully disconnecting transport resources, followed by reclaiming temporary files.
Section~\ref{sec:implementation} specifies the message payloads, transport mechanisms, and reconciliation policies that realize each phase.

\section{Protocol Specification and Implementation}
\label{sec:implementation}

This section specifies the concrete protocol artifacts that realize the AWCP framework: the message semantics governing the control plane~(\S\ref{subsec:messages}), the transport adapters provisioning the data plane~(\S\ref{subsec:transport}), and an engineering overview of the open-source reference implementation in TypeScript~(\S\ref{subsec:engineering}).

\subsection{Message Protocol}
\label{subsec:messages}

Section~\ref{subsec:lifecycle} defined five message types that drive the four-phase delegation lifecycle; here we specify their payload-level semantics.
All messages share a common header comprising a protocol version identifier, a message type discriminant, and a unique delegation identifier.
Table~\ref{tab:messages} summarizes each message type, its directionality, and primary payload.

\begin{table}[t]
\centering
\caption{AWCP message types. All messages carry a shared header with protocol version, type discriminant, and delegation identifier. D and E denote Delegator and Executor, respectively.}
\label{tab:messages}
\vspace{4pt}
\small
\begin{tabular}{@{}llll@{}}
\toprule
\textbf{Type} & \textbf{Direction} & \textbf{Phase} & \textbf{Key Payload Fields} \\
\midrule
\texttt{INVITE} & D $\to$ E & Negotiation & TaskSpec, LeaseConfig, EnvironmentDecl \\
\texttt{ACCEPT} & E $\to$ D & Negotiation & ExecutorWorkDir, ExecutorConstraints \\
\texttt{START}  & D $\to$ E & Provisioning & ActiveLease, TransportHandle \\
\texttt{DONE}   & E $\to$ D & Completion & FinalSummary, Highlights \\
\texttt{ERROR}  & Bidirectional & Any & ErrorCode, Message, Hint \\
\bottomrule
\end{tabular}
\end{table}

The \texttt{INVITE} message encapsulates the full delegation proposal: a \textit{TaskSpec} containing a natural-language description and agent prompt; a \textit{LeaseConfig} specifying the time-to-live in seconds and the access mode (read-only or read-write); and an \textit{EnvironmentDeclaration} listing the filesystem resources to be projected.
Optional fields include executor requirements and authentication credentials.

The \texttt{ACCEPT} response carries the Executor's allocated working directory and an \textit{ExecutorConstraints} structure that allows the Executor to narrow the proposed lease, for example by capping the time-to-live at a locally configured maximum or restricting the access mode to read-only.
This negotiation mechanism allows Executors to enforce local resource policies without rejecting the delegation outright.

The \texttt{START} message activates the data plane by conveying an \textit{ActiveLease} with an absolute expiration timestamp and a \textit{TransportHandle}, which is a discriminated union whose variant is determined by the selected transport type.
Each variant carries the transport-specific credentials required to establish the data channel, as detailed in \S\ref{subsec:transport}.

The \texttt{DONE} message signals successful task completion with a textual summary and optional highlights enumerating key changes.
The \texttt{ERROR} message is unique in being bidirectional: either party may emit it to signal failure, carrying a machine-readable error code, a human-readable description, and an optional hint suggesting corrective action.

\subsection{Transport Adapters}
\label{subsec:transport}

As established in \S\ref{subsec:architecture}, the transport layer is realized through pluggable adapter pairs whose concrete interfaces are detailed below.

\paragraph{Adapter interfaces.}
Rather than a single unified abstraction, the implementation defines two complementary adapter interfaces reflecting the inherently asymmetric roles in workspace delegation.
The \textit{DelegatorTransportAdapter} is responsible for packaging the local workspace into a transport-specific handle and integrating result snapshots received from the Executor.
The \textit{ExecutorTransportAdapter} handles the inverse: verifying system-level prerequisites, provisioning the workspace from the received handle, and capturing snapshots for return transmission.
Both adapters share a two-phase cleanup contract that includes \texttt{detach} for graceful disconnection followed by \texttt{release} for final resource reclamation, which mirrors the protocol's completion semantics defined in \S\ref{subsec:lifecycle}.

As introduced in \S\ref{subsec:architecture}, each adapter declares its \textit{capabilities}, particularly its support for live synchronization, which enables the control layer to select the appropriate reconciliation strategy at runtime.

\begin{table}[t]
\centering
\caption{Comparison of AWCP transport adapters across key operational dimensions.}
\label{tab:transports}
\vspace{4pt}
\small
\begin{tabular}{@{}lllll@{}}
\toprule
 & \textbf{SSHFS} & \textbf{Archive} & \textbf{Storage} & \textbf{Git} \\
\midrule
\textbf{Mechanism}      & SSH + FUSE mount & HTTP + ZIP    & Pre-signed URLs   & Branch-based VCS \\
\textbf{Live sync}      & Yes              & No            & No                & No \\
\textbf{Snapshots}      & N/A (live)       & Yes           & Yes               & Yes \\
\textbf{Data integrity} & SSH channel      & SHA-256       & SHA-256           & Git SHA \\
\textbf{Best suited for} & Interactive,    & Small         & Large files,      & Auditable, \\
                         & low-latency      & workspaces    & cloud-native      & versioned \\
\bottomrule
\end{tabular}
\end{table}

\paragraph{Native transport implementations.}
Table~\ref{tab:transports} compares the four transport adapters along key operational dimensions.

The \textit{SSHFS} transport projects the Delegator's workspace via a FUSE-based mount~\citep{vangoor2017fuse} over an SSH tunnel with ephemeral credentials, providing bidirectional real-time synchronization ideal for interactive tasks.
The \textit{Archive} transport serializes the workspace into a base64-encoded ZIP archive transmitted inline within the \texttt{START} message, offering a self-contained mechanism that requires no external infrastructure.
The \textit{Storage} transport decouples data transfer from the control channel via pre-signed URLs, accommodating cloud-native deployments where large workspaces can be exchanged through object storage services.
The \textit{Git} transport leverages branch-based version control, with each snapshot corresponding to a commit that provides a full audit trail of workspace modifications.
As discussed in \S\ref{sec:related}, recent Git-based coordination systems~\citep{huang2025evogit, li2025agentgit} independently validate this design; AWCP's adapter generalizes the pattern into a transport-agnostic component of the delegation lifecycle.

\paragraph{Snapshot reconciliation.}
For transports that do not support live synchronization, the Executor periodically captures workspace snapshots, which are compressed archives of the current file state, and subsequently transmits them to the Delegator via SSE events.
The Delegator processes these snapshots according to a configurable \textit{snapshot policy} with three modes: \texttt{auto}, which applies incoming snapshots immediately; \texttt{staged}, which queues snapshots for explicit approval by the Delegator agent or a human operator; and \texttt{discard}, which ignores snapshots entirely for read-only delegations.
When a transport reports live synchronization capability, the policy is overridden to \texttt{auto}, as the mounted filesystem already reflects the latest state.

\subsection{Engineering Overview}
\label{subsec:engineering}

The reference implementation is structured as an npm workspace monorepo comprising eight packages organized by the dependency hierarchy shown in Table~\ref{tab:implementation}.
The \texttt{@awcp/core} package contains the protocol types, dual state machines, and error definitions with zero external dependencies, ensuring that the protocol specification remains portable.
The \texttt{@awcp/sdk} builds upon the core package to provide the \textit{DelegatorService} and \textit{ExecutorService} with JSON-based state persistence, enabling crash recovery across sessions.
The four transport packages each export a \textit{DelegatorTransport} and \textit{ExecutorTransport} pair conforming to the adapter interfaces described in \S\ref{subsec:transport}.
The \texttt{@awcp/mcp} package exposes seven MCP tools covering the full delegation lifecycle, including initiation, output monitoring, cancellation, snapshot management, and session recovery, enabling MCP-compatible agents such as Claude Desktop and Cline to manage delegations through standard tool calls.
A standalone Bun-based skill module provides equivalent functionality for chat-oriented platforms.

\begin{table}[t]
\centering
\caption{Reference implementation overview. Source lines of code~(SLoC) are counted for TypeScript files excluding tests. All packages are published under synchronized versioning.}
\label{tab:implementation}
\vspace{4pt}
\small
\begin{tabular}{@{}llrrl@{}}
\toprule
\textbf{Package} & \textbf{Role} & \textbf{Source} & \textbf{Tests} & \textbf{Dependencies} \\
\midrule
\texttt{@awcp/core}               & Protocol types, state machines   & 1,075 &  393 & --- \\
\texttt{@awcp/sdk}                & Services, persistence            & 3,643 &  635 & core \\
\texttt{@awcp/mcp}                & 7 MCP tools                      & 1,666 &  286 & sdk, transports \\
\texttt{@awcp/transport-sshfs}    & SSH + FUSE mount                 & 1,052 &  447 & core \\
\texttt{@awcp/transport-archive}  & HTTP + ZIP                       &   331 &  327 & core \\
\texttt{@awcp/transport-storage}  & Pre-signed URLs                  &   364 &  348 & core, archive \\
\texttt{@awcp/transport-git}      & Branch-based VCS                 &   328 &  141 & core, archive \\
\texttt{awcp-skill}               & Bun CLI skill module             &   769 &  --- & --- \\
\midrule
\textbf{Total}                    &                                  & \textbf{9,228} & \textbf{2,577} & \\
\bottomrule
\end{tabular}
\end{table}

The codebase totals approximately 9,200 lines of TypeScript source code with 2,600 lines of tests organized into 161 test cases across 12 test files, executed via the Vitest framework.
All packages enforce TypeScript strict mode and are published under synchronized versioning.

\section{Demonstration}
\label{sec:demo}

To validate AWCP in practical settings, we demonstrate two live collaboration scenarios in which agents with complementary capabilities cooperate through workspace delegation.
The scenarios are selected to exercise distinct dimensions of asymmetry, including capability, trust, and domain knowledge, across different integration modes and transport mechanisms.

\subsection{Cross-Modal Dataset Curation}

\begin{figure}[t]
    \centering
    \includegraphics[width=0.88\textwidth, trim=0pt 0pt 0pt 0pt, clip]{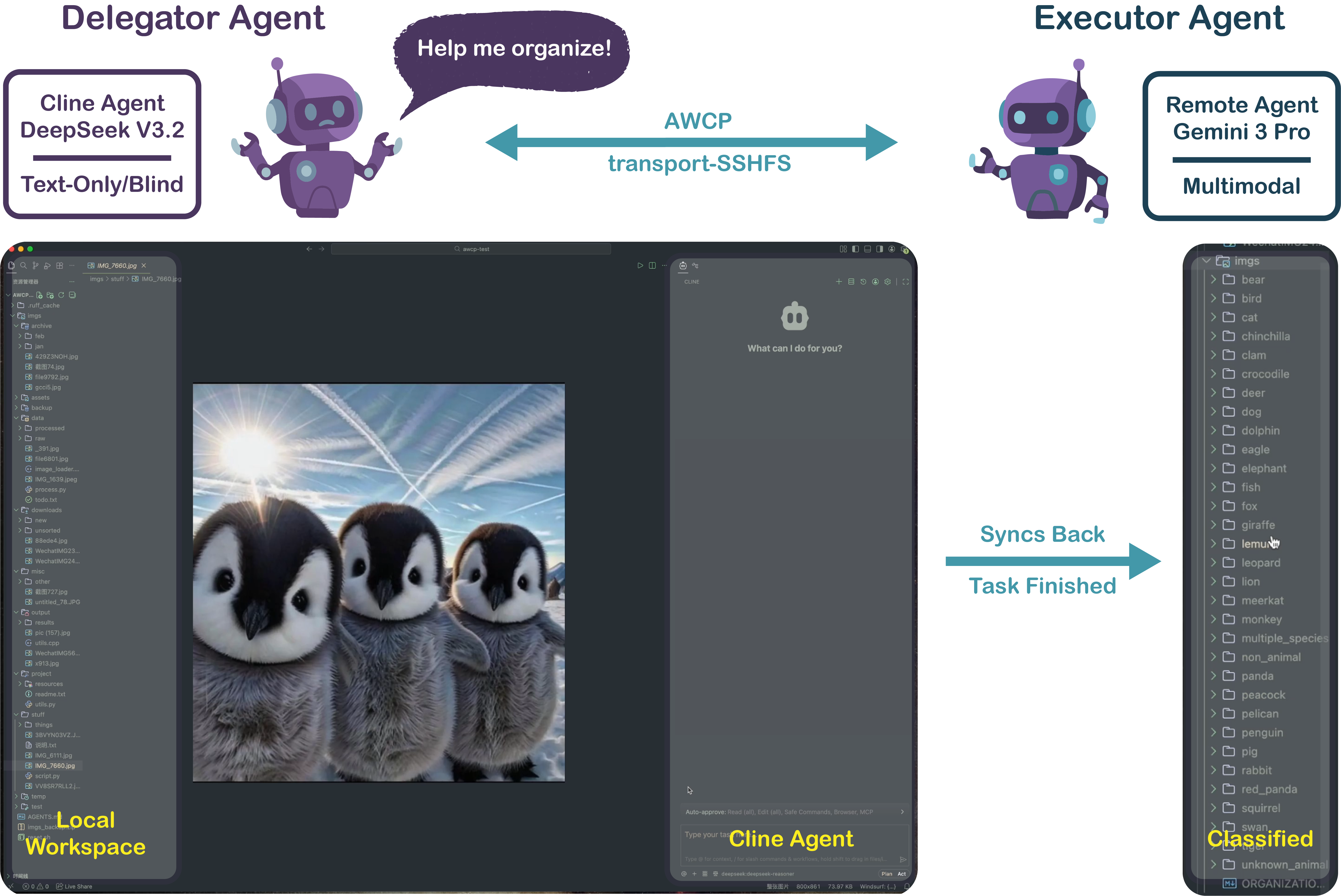}
    \caption{Cross-modal dataset curation via SSHFS transport. A text-only Delegator~(Cline + DeepSeek V3.2) delegates a cluttered image directory to a multimodal Executor~(Gemini 3 Pro). The Executor classifies, filters, and reorganizes files; changes synchronize bidirectionally in real time through the FUSE mount.}
    \label{fig:demo1}
\end{figure}

As illustrated in Figure~\ref{fig:demo1}, a text-only coding agent is tasked with organizing a cluttered directory containing over 100 images mixed with corrupted files, empty placeholders, and non-image artifacts scattered across deeply nested subdirectories.
The task requires classifying images by content, removing corrupted files, and reorganizing the directory semantically, all of which demand visual perception that the text-only Delegator lacks.

We configure VS Code with the Cline extension as the Delegator~\citep{cline2024}, running DeepSeek V3.2 as its backbone LLM, and integrate AWCP through its MCP server.
The Executor runs Gemini 3 Pro, a multimodal model capable of visual understanding.
The SSHFS transport is selected for this scenario, as the interactive nature of the task, which requires the Executor to iteratively traverse directories, inspect individual files, and reorganize content, benefits significantly from live bidirectional synchronization.

Upon receiving the delegation, the Executor mounts the Delegator's workspace as a local directory via FUSE.
The remote agent then analyzes each file using its vision capabilities, identifies corrupted images and non-image artifacts for removal, and reorganizes valid images into semantically labeled subdirectories~(e.g., \texttt{bear/}, \texttt{penguin/}, \texttt{giraffe/}).
Every file operation, including creation, deletion, renaming, synchronizes in real time to the Delegator's local filesystem through the live mount.
Upon completion, the Delegator observes a clean, categorized directory structure without having performed any manual file transfer.

This scenario demonstrates \textit{capability asymmetry}, in which the text-only Delegator lacks visual perception but gains access to multimodal analysis through workspace delegation, as well as \textit{knowledge asymmetry}, since the Executor brings domain-specific image classification expertise that the Delegator does not possess.
Throughout, the Executor operates directly on the delegated files as if they were local, applying its own toolchain without any manual file transfer by the Delegator.

\subsection{Multi-Round Compliance Stamping}

\begin{figure}[t]
    \centering
    \includegraphics[width=0.88\textwidth, trim=10pt 10pt 10pt 10pt, clip]{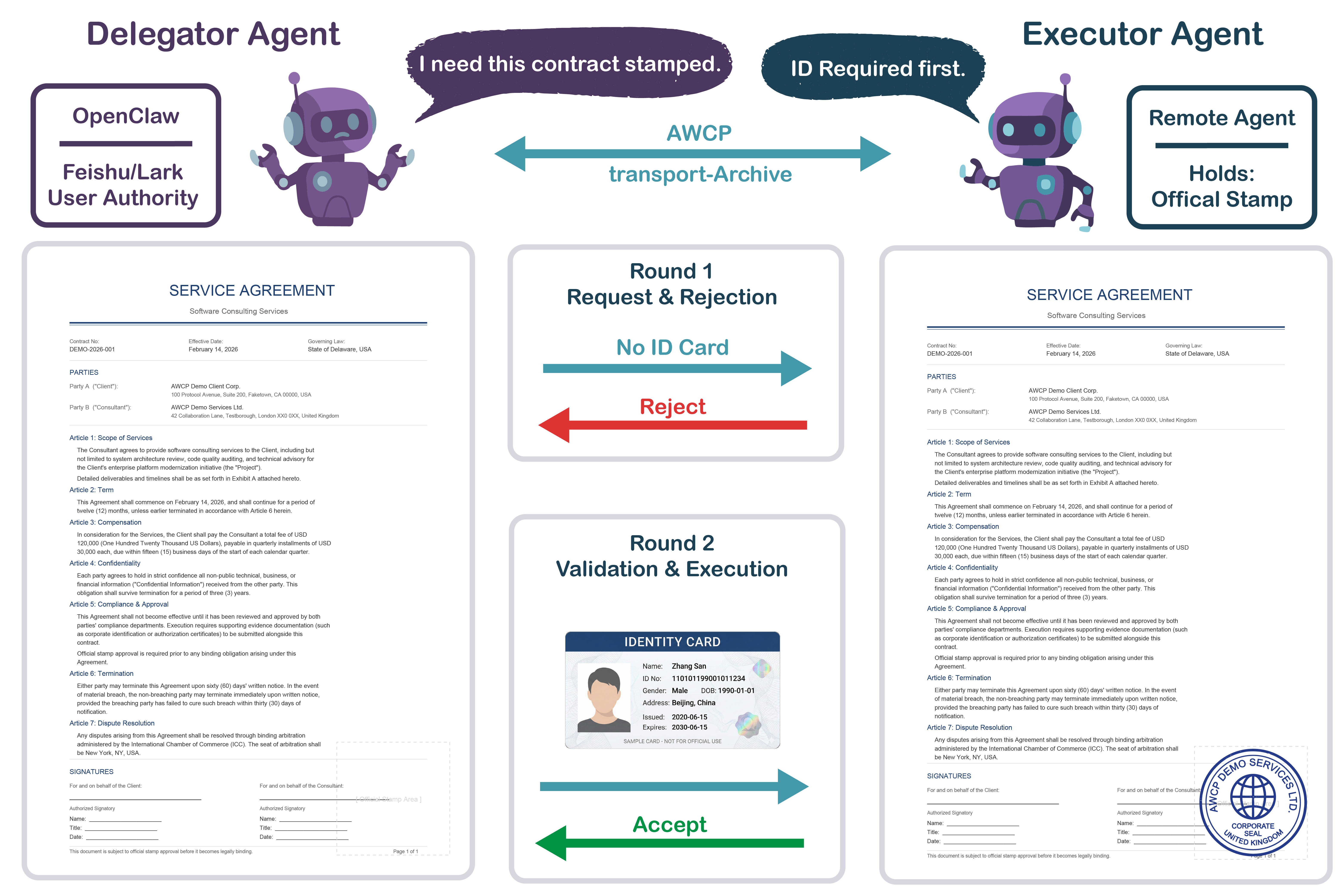}
    \caption{Multi-round compliance stamping via Archive transport. A Delegator~(OpenClaw on Feishu) delegates a contract to a Compliance Auditor Executor for official stamping. Round~1 is rejected for missing identity verification; Round~2 succeeds after the user supplies the required document, demonstrating AWCP's support for iterative delegation workflows.}
    \label{fig:demo2}
\end{figure}

As shown in Figure~\ref{fig:demo2}, a user requests official contract stamping via an OpenClaw agent on the Feishu platform.
Since the local agent lacks authorization for the digital seal, it utilizes an AWCP skill module to discover a remote Compliance Auditor and delegates a workspace containing \texttt{contract.pdf}.
The Archive transport is selected here: the workspace consists of a single PDF file, making the self-contained ZIP-based transfer more practical than establishing a persistent SSHFS mount.

The delegation proceeds through two rounds, each constituting a complete AWCP lifecycle.
In Round~1, the Executor inspects the submitted materials, identifies a missing identity verification document, and returns a structured rejection via the \texttt{ERROR} message, at which point the Delegator transitions to the \texttt{error} terminal state and the Executor releases its resources.
Upon receiving the required identity card image from the user, the Delegator agent initiates a fresh delegation~(Round~2) with the updated workspace.
The Executor verifies completeness, passes the compliance audit, applies the official stamp, and returns \texttt{contract\_signed.pdf} as a snapshot.
The Delegator applies the snapshot to obtain the stamped document.

This scenario demonstrates both \textit{trust asymmetry}, where the privileged stamping operation executes exclusively on the authorized Executor without exposing the digital seal to the Delegator environment~\citep{wu2025isolategpt}, and \textit{knowledge asymmetry}, as the Compliance Auditor possesses domain-specific regulatory expertise that the general-purpose Delegator agent lacks.
Furthermore, the two-round workflow illustrates the protocol's native support for \textit{iterative workflows}, where each round is a self-contained delegation with independent state machine instances.

\section{Conclusion}
\label{sec:conclusion}

We have presented AWCP, a protocol that formalizes inter-agent collaboration through temporary workspace delegation, grounding interaction in the filesystem rather than in message exchange.
Three contributions emerge from this work: a workspace-centric collaboration paradigm that moves beyond payload serialization; a decoupled architecture whose pluggable transports adapt to diverse network conditions; and open-source infrastructure demonstrating practical feasibility through live asymmetric collaboration.

Our demonstrations confirm several properties of this design.
The protocol accommodates diverse integration modes, from MCP tool servers for IDE-based agents to skill modules for chat-oriented platforms.
The pluggable transport layer adapts to task characteristics, selecting SSHFS for interactive operations requiring live synchronization and Archive for lightweight document exchange.
The dual state machine formalization supports both single-round and multi-round collaboration patterns with independent lifecycle governance.
In both scenarios, Executors work directly on delegated files as though they were local, confirming that workspace projection eliminates the context loss inherent in message-based coordination.

Several directions remain open for future work.
On the transport side, peer-to-peer and edge-optimized adapters would extend AWCP to bandwidth-constrained and decentralized deployments~\citep{zhang2026toward}, and CRDT-based synchronization mechanisms~\citep{pugachev2025codecrdt} could enable conflict-free concurrent workspace access for multi-party delegations.
On the governance side, granular access control policies, including file-level permissions, role-based authorization, and audit logging, constitute essential prerequisites for multi-party delegations involving sensitive workspaces~\citep{yang2025survey}.
On the scale side, extending the current one-to-one delegation model to federated multi-agent coalitions, where multiple Executors collaborate on partitioned workspaces under incentive-compatible coordination~\citep{yang2026internetagentic, wooldridge2025fetchai}, would align AWCP with the emerging Internet of Agentic AI.
On the discovery side, integrating cross-boundary capability advertisement would allow Delegators to locate suitable Executors dynamically, closing the gap between workspace delegation and the broader agent coordination stack~\citep{giusti2025federation}.
As agents grow more specialized, a standardized workspace layer will become essential infrastructure for the Agentic Web, letting capable but isolated agents finally collaborate at the depth their tasks demand.

\section*{Acknowledgements}

This research was supported by National Natural Science Foundation of China (62322603 and 625B2185). The work is accomplished at Shanghai Innovation Institute.

\appendix

\bibliographystyle{unsrtnat}
\bibliography{main}

\end{document}

%% file: figs/protocol-landscape.tex
%%%%%%%%%%%%%%%%%%%%%%%%%%%%%%%%%%%%%%%%%%%%%%%%%%%%%%%%%%%%%%%%%%%%%%%%%%%%%%%
% AWCP Protocol Landscape - Relationship with MCP, A2A, and other protocols
%
% Required: \usepackage{tikz}
%           \usetikzlibrary{positioning, arrows.meta, fit, backgrounds, calc}
%%%%%%%%%%%%%%%%%%%%%%%%%%%%%%%%%%%%%%%%%%%%%%%%%%%%%%%%%%%%%%%%%%%%%%%%%%%%%%%

\begin{figure}[t]
\centering
\resizebox{0.92\textwidth}{!}{%
\begin{tikzpicture}[
    % === Node styles ===
    agent/.style={rectangle, rounded corners=6pt, draw=black!35, thick,
        fill=green!5, minimum width=2.0cm, minimum height=1.3cm,
        font=\small\bfseries, align=center},
    svc/.style={rectangle, rounded corners=5pt, draw=black!45, thick,
        fill=blue!6, minimum width=2.6cm, minimum height=1.3cm,
        font=\small\bfseries, align=center},
    pill/.style={rectangle, rounded corners=3pt, draw=black!20,
        inner sep=4pt, font=\scriptsize\bfseries, text=black!65},
    % === Arrow styles ===
    arr/.style={-{Stealth[length=2.5mm, width=2mm]}, thick, black!55},
    % === Label styles ===
    desc/.style={font=\scriptsize, text=black!40, align=center}
]

% =====================================================================
% Left column: User + Delegator Agent
% =====================================================================
\node[agent, fill=yellow!6, draw=black!30] (user) at (0, 0.4) {User};
\node[agent] (dagent) at (0, -2.0) {Delegator\\Agent};
\draw[arr] (user) -- (dagent);

% =====================================================================
% Center: AWCP boundary
% =====================================================================

% --- Services ---
\node[svc] (dserv) at (6.2, -0.2) {Delegator\\Service};
\node[svc] (eserv) at (11.4, -0.2) {Executor\\Service};

% --- Control plane: bidirectional HTTP/SSE between services ---
\draw[{Stealth[length=2.5mm, width=2mm]}-{Stealth[length=2.5mm, width=2mm]},
    thick, black!40, densely dashed] (dserv.east) -- (eserv.west);
\node[desc] at (8.8, 0.25) {HTTP / SSE};

% --- Transport plane (below services) ---
\node[pill, fill=orange!10, draw=orange!30, text=orange!70!black]
    (trans) at (8.8, -2.2)
    {SSHFS~~$\vert$~~Archive~~$\vert$~~Storage~~$\vert$~~Git};
\node[desc] at (8.8, -2.75) {Workspace Transport};
% --- Transport connection: bus-style vertical stubs + horizontal bar ---
\coordinate (busY) at (0, -1.25);  % y-level of horizontal bus
\draw[orange!55, line width=1pt]
    (dserv.south) -- (dserv.south |- busY);
\draw[orange!55, line width=1pt]
    (eserv.south) -- (eserv.south |- busY);
\draw[orange!55, line width=1pt]
    (dserv.south |- busY) -- (eserv.south |- busY);
\draw[arr, orange!55, line width=1pt]
    (8.8, -1.25) -- (trans.north);

% --- AWCP boundary box (enlarged, label outside) ---
\begin{scope}[on background layer]
    \draw[rounded corners=10pt, draw=orange!40, thick,
        fill=orange!3, densely dashed]
        (4.3, 1.2) rectangle (13.3, -3.3);
\end{scope}
\node[font=\footnotesize\bfseries, text=orange!60!black, anchor=south west]
    at (4.45, 1.25) {AWCP};

% =====================================================================
% Right column: Executor Agent
% =====================================================================
\node[agent] (eagent) at (17.6, -2.0) {Executor\\Agent};

% =====================================================================
% Connections: Agents ↔ AWCP services
% =====================================================================

% --- Delegator Agent → Delegator Service ---
\draw[arr] (dagent.east) -- ++(1.6, 0) |- (dserv.west);

% Integration pill: positioned in the gap, clearly left of AWCP border
\node[pill, fill=purple!6] at (2.5, -1.2)
    {MCP~~$\vert$~~Skill~~$\vert$~~Tools};
\node[desc] at (2.5, -1.7) {Integration Interfaces};

% --- Executor Service → Executor Agent ---
\draw[arr] (eserv.east) -- ++(1.6, 0) |- (eagent.west);

% Adapter pill: positioned in the gap, clearly right of AWCP border
\node[pill, fill=purple!6] at (15.1, -1.2)
    {A2A~~$\vert$~~ANP~~$\vert$~~HTTP};
\node[desc] at (15.1, -1.7) {Adapter Layer};

\end{tikzpicture}%
}
\caption{AWCP in the agentic protocol landscape. Delegator agents access AWCP through integration interfaces such as MCP tool servers or skill modules. The control plane uses standard HTTP and Server-Sent Events for signaling between AWCP services. Executor agents are invoked through an extensible adapter layer supporting A2A, ANP, or direct HTTP. The transport plane provides pluggable filesystem-level access that complements message-level coordination.}
\label{fig:protocol-landscape}
\end{figure}

%% file: figs/awcp-sequence-diagram.tex
%%%%%%%%%%%%%%%%%%%%%%%%%%%%%%%%%%%%%%%%%%%%%%%%%%%%%%%%%%%%%%%%%%%%%%%%%%%%%%%
% AWCP Protocol Sequence Diagram - Professional UML 2.0 Style
% 
% Required: \usepackage{tikz}
%           \usetikzlibrary{positioning, arrows.meta, fit, backgrounds, calc}
%%%%%%%%%%%%%%%%%%%%%%%%%%%%%%%%%%%%%%%%%%%%%%%%%%%%%%%%%%%%%%%%%%%%%%%%%%%%%%%

\begin{figure}[!htb]
\centering
\resizebox{0.6\textwidth}{!}{%
\begin{tikzpicture}[
    % === Core styles ===
    participant/.style={rectangle, draw=black, thick, fill=white,
        minimum width=2.2cm, minimum height=0.7cm, font=\small\bfseries},
    lifeline/.style={dashed, gray!60},
    % === Message arrows ===
    msg/.style={-{Stealth[length=2mm, width=1.5mm]}, thick},
    ret/.style={-{Stealth[length=2mm, width=1.5mm]}, thick, dashed},
    sse/.style={-{Stealth[length=2mm, width=1.5mm]}, thick, blue!70},
    % === Labels ===
    mlabel/.style={font=\scriptsize\ttfamily, fill=white, inner sep=1.5pt},
    plabel/.style={font=\scriptsize\itshape, text=black!70},
    % === Fragment (UML combined fragment) ===
    fragment/.style={draw=black!60, fill=white},
    fraglabel/.style={font=\scriptsize\bfseries, fill=black!10, draw=black!60, 
        inner sep=2pt, anchor=north west}
]

% ========== Participants ==========
\node[participant] (D) at (0, 0) {Delegator};
\node[participant] (E) at (6, 0) {Executor};

% ========== Lifelines ==========
\draw[lifeline] (0, -0.35) -- (0, -13.02);
\draw[lifeline] (6, -0.35) -- (6, -13.02);

%==========================================================================
% PHASE 1: Negotiation
%==========================================================================
\begin{scope}[on background layer]
  \draw[fragment] (-1.2, -0.6) rectangle (7.2, -2.8);
\end{scope}
\node[fraglabel] at (-1.2, -0.6) {Negotiation};

% INVITE
\draw[msg] (0, -1.2) -- node[mlabel, above] {INVITE} (6, -1.2);
\node[plabel] at (3, -1.5) {task, lease, environment};

% ACCEPT (response)
\draw[ret] (6, -2.2) -- node[mlabel, above] {ACCEPT} (0, -2.2);
\node[plabel] at (3, -2.5) {workDir, constraints};

%==========================================================================
% PHASE 2: Provisioning  
%==========================================================================
\begin{scope}[on background layer]
  \draw[fragment] (-1.2, -3.0) rectangle (7.2, -5.8);
\end{scope}
\node[fraglabel] at (-1.2, -3.0) {Provisioning};

% Self-call: Prepare Transport (proper UML self-message)
\draw[msg] (0, -3.6) -- ++(0.8, 0) -- ++(0, -0.3) -- ++(-0.8, 0);
\node[mlabel, anchor=west] at (0.9, -3.7) {Prepare Transport};

% START
\draw[msg] (0, -4.5) -- node[mlabel, above] {START} (6, -4.5);
\node[plabel] at (3, -4.8) {lease, workDir (credentials)};

% {ok} response
\draw[ret] (6, -5.5) -- node[mlabel, above] {\{ok: true\}} (0, -5.5);

%==========================================================================
% PHASE 3: Execution (SSE Stream)
%==========================================================================
\begin{scope}[on background layer]
  \draw[fragment] (-1.2, -6.0) rectangle (7.2, -11.1);
\end{scope}
\node[fraglabel] at (-1.2, -6.0) {Execution};

% Self-call: Setup Workspace
\draw[msg] (6, -6.3) -- ++(-0.8, 0) -- ++(0, -0.3) -- ++(0.8, 0);
\node[mlabel, anchor=east] at (5.1, -6.4) {Setup Workspace};

% SSE subscription
\draw[sse] (0, -7.2) -- node[mlabel, above] {SSE: /tasks/:taskId/events} (6, -7.2);

% status event
\draw[sse] (6, -7.9) -- node[mlabel, above] {event: status} (0, -7.9);
\node[plabel] at (3, -8.2) {running};

% Self-call: Execute Task
\draw[msg] (6, -8.7) -- ++(-0.8, 0) -- ++(0, -0.3) -- ++(0.8, 0);
\node[mlabel, anchor=east] at (5.1, -8.8) {Execute Task};

% Self-call: Teardown
\draw[msg] (6, -9.5) -- ++(-0.8, 0) -- ++(0, -0.3) -- ++(0.8, 0);
\node[mlabel, anchor=east] at (5.1, -9.6) {Teardown};

% snapshot event
\draw[sse] (6, -10.2) -- node[mlabel, above] {event: snapshot} (0, -10.2);
\node[plabel] at (3, -10.5) {snapshotId, data, recommended};

% done event
\draw[sse] (6, -10.9) -- node[mlabel, above] {event: done} (0, -10.9);

%==========================================================================
% PHASE 4: Completion
%==========================================================================
\begin{scope}[on background layer]
  \draw[fragment] (-1.2, -11.3) rectangle (7.2, -13.0);
\end{scope}
\node[fraglabel] at (-1.2, -11.3) {Completion};

% Self-call: Apply Snapshot
\draw[msg] (0, -11.9) -- ++(0.8, 0) -- ++(0, -0.3) -- ++(-0.8, 0);
\node[mlabel, anchor=west] at (0.9, -12.0) {Apply Snapshot};

% ACK
\draw[msg] (0, -12.7) -- node[mlabel, above] {ACK} (6, -12.7);

\end{tikzpicture}%
}
\caption{AWCP four-phase message sequence between Delegator and Executor.
Solid, dashed, and blue arrows denote synchronous HTTP requests, responses, and asynchronous SSE events, respectively.}
\label{fig:awcp-protocol}
\end{figure}

%% file: figs/state-machines.tex
%%%%%%%%%%%%%%%%%%%%%%%%%%%%%%%%%%%%%%%%%%%%%%%%%%%%%%%%%%%%%%%%%%%%%%%%%%%%%%%
% AWCP Dual State Machines - Delegation (Delegator) and Assignment (Executor)
%
% Required: \usepackage{tikz}
%           \usetikzlibrary{positioning, arrows.meta, fit, backgrounds, calc}
%%%%%%%%%%%%%%%%%%%%%%%%%%%%%%%%%%%%%%%%%%%%%%%%%%%%%%%%%%%%%%%%%%%%%%%%%%%%%%%

\begin{figure}[t]
\centering
\resizebox{0.88\textwidth}{!}{%
\begin{tikzpicture}[
    % === State styles ===
    state/.style={rectangle, rounded corners=5pt, draw=black!60, thick,
        fill=white, minimum width=1.8cm, minimum height=0.75cm,
        font=\small, align=center},
    init/.style={state, fill=blue!6},
    terminal/.style={state, fill=black!8, draw=black!40, text=black!65},
    % === Arrow styles ===
    trans/.style={-{Stealth[length=2.2mm, width=1.8mm]}, thick, black!65},
    sync/.style={-{Stealth[length=2.2mm, width=1.8mm]}, thick, blue!55,
        densely dashed},
    % === Label styles ===
    evlabel/.style={font=\scriptsize\ttfamily, fill=white, inner sep=1.5pt,
        text=black!70},
    synclabel/.style={font=\scriptsize\itshape, fill=white, inner sep=1.5pt,
        text=blue!60},
    fraglabel/.style={font=\small\bfseries, text=black!60},
    % === Fragment ===
    fragment/.style={draw=black!20, rounded corners=8pt, fill=black!2}
]

% === Vertical spacing ===
\def\ystep{1.55}

%==========================================================================
% LEFT: DelegationStateMachine (Delegator side)
%==========================================================================

% Happy-path states — vertical spine
\node[init]     (d-created)   at (0, 0)              {\texttt{created}};
\node[state]    (d-invited)   at (0, -1*\ystep)      {\texttt{invited}};
\node[state]    (d-accepted)  at (0, -2*\ystep)      {\texttt{accepted}};
\node[state]    (d-started)   at (0, -3*\ystep)      {\texttt{started}};
\node[state]    (d-running)   at (0, -4*\ystep)      {\texttt{running}};
\node[terminal] (d-completed) at (0, -5*\ystep)      {\texttt{completed}};

% Happy-path transitions
\draw[trans] (d-created)  -- node[evlabel, right, xshift=1pt] {SEND\_INVITE}    (d-invited);
\draw[trans] (d-invited)  -- node[evlabel, right, xshift=1pt] {RECV\_ACCEPT}    (d-accepted);
\draw[trans] (d-accepted) -- node[evlabel, right, xshift=1pt] {SEND\_START}     (d-started);
\draw[trans] (d-started)  -- node[evlabel, right, xshift=1pt] {SETUP\_COMPLETE} (d-running);
\draw[trans] (d-running)  -- node[evlabel, right, xshift=1pt] {RECV\_DONE}      (d-completed);

% Terminal states — grouped to the left, generous vertical spacing
\node[terminal] (d-error)     at (-2.6, -5*\ystep)       {\texttt{error}};
\node[terminal] (d-cancelled) at (-2.6, -5*\ystep-1.1)   {\texttt{cancelled}};
\node[terminal] (d-expired)   at (-2.6, -5*\ystep-2.2)   {\texttt{expired}};

% Single aggregate arrow for exceptional transitions
\draw[-{Stealth[length=2.2mm, width=1.8mm]}, thick, red!40, dashed]
    (-1.15, -0.5*\ystep) -- (-1.15, -4.5*\ystep) -- (d-error.north east);
\node[font=\scriptsize, text=red!45, align=left, anchor=east] 
    at (-1.25, -2.5*\ystep) {\textit{any non-terminal}};

% Annotation for cancelled/expired
\node[font=\scriptsize, text=black!40, align=center, anchor=north] 
    at (-2.6, -5*\ystep-3.1) {via \textsc{error} / \textsc{cancel} / \textsc{expire}};

% Fragment box
\begin{scope}[on background layer]
    \draw[fragment] (-4.05, 1.55) rectangle (1.85, -5*\ystep-3.85);
\end{scope}
\node[fraglabel] at (-1.1, 1.2) {DelegationStateMachine};
\node[font=\scriptsize, text=black!40] at (-1.1, 0.8) {(Delegator side)};

%==========================================================================
% RIGHT: AssignmentStateMachine (Executor side)
%==========================================================================

% Align sync points: pending at invited level, active at started level,
% completed at running level
\def\xr{6.2}

\node[init]     (e-pending)   at (\xr, -1*\ystep)   {\texttt{pending}};
\node[state]    (e-active)    at (\xr, -3*\ystep)    {\texttt{active}};
\node[terminal] (e-completed) at (\xr, -5*\ystep)    {\texttt{completed}};
\node[terminal] (e-error)     at ({\xr+2.2}, -5*\ystep) {\texttt{error}};

% Happy-path transitions
\draw[trans] (e-pending) -- node[evlabel, left, xshift=-1pt] {RECV\_START} (e-active);
\draw[trans] (e-active)  -- node[evlabel, left, xshift=-1pt] {TASK\_COMPLETE} (e-completed);

% Error transitions — single aggregate arrow
\draw[-{Stealth[length=2.2mm, width=1.8mm]}, thick, red!40, dashed]
    ({\xr+1.15}, -1.5*\ystep) -- ({\xr+1.15}, -4.5*\ystep) -- (e-error.north west);
\node[font=\scriptsize, text=red!45, align=right, anchor=west]
    at ({\xr+1.25}, -3*\ystep) {\textit{any non-}\\\textit{terminal}};

% Annotation
\node[font=\scriptsize, text=black!40, align=center, anchor=north]
    at ({\xr+2.2}, -5*\ystep-0.7) {via \textsc{error} / \textsc{cancel}};

% Fragment box
\begin{scope}[on background layer]
    \draw[fragment] ({\xr-1.75}, 1.55) rectangle ({\xr+3.55}, -5*\ystep-1.4);
\end{scope}
\node[fraglabel] at ({\xr+0.9}, 1.2) {AssignmentStateMachine};
\node[font=\scriptsize, text=black!40] at ({\xr+0.9}, 0.8) {(Executor side)};

%==========================================================================
% Synchronization arrows (horizontal, between aligned states)
%==========================================================================

% ACCEPT: Executor sends ACCEPT → Delegator transitions invited→accepted,
%         Executor creates assignment in pending
\draw[sync] (d-invited.east) ++(0.15,0) -- 
    node[synclabel, above] {\textnormal{\texttt{ACCEPT}}} 
    ([xshift=-0.15cm]e-pending.west);

% START: Delegator sends START → triggers pending→active
\draw[sync] (d-accepted.east) ++(0.15,0) -- ++(0.6,0) |- 
    node[synclabel, above, pos=0.75] {\textnormal{\texttt{START}}} 
    ([xshift=-0.15cm]e-active.west);

% DONE: Executor sends DONE → Delegator transitions running→completed
\draw[{Stealth[length=2.2mm, width=1.8mm]}-, thick, blue!55, densely dashed] 
    (d-running.east) ++(0.15,0) -- ++(0.6,0) |- 
    node[synclabel, above, pos=0.75] {\textnormal{\texttt{DONE}}} 
    ([xshift=-0.15cm]e-completed.west);

\end{tikzpicture}%
}
\caption{Dual state machines governing the AWCP delegation lifecycle. The \textit{DelegationStateMachine}~(left) tracks nine states on the Delegator side; the \textit{AssignmentStateMachine}~(right) tracks four states on the Executor side. Solid arrows trace the happy path; a single dashed red path indicates that any non-terminal state may transition to a terminal state upon error, cancellation, or lease expiration. Blue dashed arrows mark cross-machine synchronization via protocol messages.}
\label{fig:state-machines}
\end{figure}

%% file: main.bbl
\begin{thebibliography}{42}
\providecommand{\natexlab}[1]{#1}
\providecommand{\url}[1]{\texttt{#1}}
\expandafter\ifx\csname urlstyle\endcsname\relax
  \providecommand{\doi}[1]{doi: #1}\else
  \providecommand{\doi}{doi: \begingroup \urlstyle{rm}\Url}\fi

\bibitem[Wang et~al.(2024)Wang, Ma, Feng, Zhang, Yang, Zhang, Chen, Tang, Chen, Lin, Zhao, Wei, and Wen]{wang2024survey}
Lei Wang, Chen Ma, Xueyang Feng, Zeyu Zhang, Hao Yang, Jingsen Zhang, Zhiyuan Chen, Jiakai Tang, Xu~Chen, Yankai Lin, Wayne~Xin Zhao, Zhewei Wei, and Ji-Rong Wen.
\newblock A survey on large language model based autonomous agents.
\newblock \emph{Frontiers of Computer Science}, 18\penalty0 (6):\penalty0 186345, 2024.
\newblock \doi{10.1007/s11704-024-40231-1}.

\bibitem[Yang et~al.(2025{\natexlab{a}})Yang, Ma, Huang, Chai, Gong, Geng, Zhou, Wen, Fang, Chen, et~al.]{yang2025agenticweb}
Yingxuan Yang, Mulei Ma, Yuxuan Huang, Huacan Chai, Chenyu Gong, Haoran Geng, Yuanjian Zhou, Ying Wen, Meng Fang, Muhao Chen, et~al.
\newblock Agentic web: Weaving the next web with {AI} agents.
\newblock \emph{arXiv preprint arXiv:2507.21206}, 2025{\natexlab{a}}.

\bibitem[Ehtesham et~al.(2025)Ehtesham, Singh, Gupta, and Kumar]{ehtesham2025survey}
Abul Ehtesham, Aditi Singh, Gaurav~Kumar Gupta, and Saket Kumar.
\newblock A survey of agent interoperability protocols: {Model Context Protocol} ({MCP}), {Agent Communication Protocol} ({ACP}), {Agent-to-Agent Protocol} ({A2A}), and {Agent Network Protocol} ({ANP}).
\newblock \emph{arXiv preprint arXiv:2505.02279}, 2025.

\bibitem[Cemri et~al.(2025)Cemri, Pan, Yang, Agrawal, Chopra, Tiwari, Keutzer, Parameswaran, Klein, Ramchandran, Zaharia, Gonzalez, and Stoica]{cemri2025mast}
Mert Cemri, Melissa~Z. Pan, Shuyi Yang, Lakshya~A. Agrawal, Bhavya Chopra, Rishabh Tiwari, Kurt Keutzer, Aditya Parameswaran, Dan Klein, Kannan Ramchandran, Matei Zaharia, Joseph~E. Gonzalez, and Ion Stoica.
\newblock Why do multi-agent {LLM} systems fail?
\newblock \emph{arXiv preprint arXiv:2503.13657}, 2025.

\bibitem[{Anthropic}(2024)]{anthropic2024mcp}
{Anthropic}.
\newblock Model context protocol specification.
\newblock \url{https://modelcontextprotocol.io/specification}, 2024.
\newblock Accessed: 2026-02-14.

\bibitem[{Google}(2025)]{google2025a2a}
{Google}.
\newblock Agent2agent ({A2A}) protocol specification.
\newblock \url{https://a2a-protocol.org/latest/specification/}, 2025.
\newblock Accessed: 2026-02-14.

\bibitem[Nie et~al.(2026)Nie, Guo, Cui, Yang, Chen, De, Zhang, Liao, Huang, Yang, Han, Peng, Chen, Tang, Liu, Zhou, Hu, Tang, Lin, Liu, Wen, Zhou, and Zhang]{holos2026}
Xiaohang Nie, Zihan Guo, Zicai Cui, Jiachi Yang, Zeyi Chen, Leheyi De, Yu~Zhang, Junwei Liao, Bo~Huang, Yingxuan Yang, Zhi Han, Zimian Peng, Linyao Chen, Wenzheng~Tom Tang, Zongkai Liu, Tao Zhou, Botao~Amber Hu, Shuyang Tang, Jianghao Lin, Weiwen Liu, Muning Wen, Yuanjian Zhou, and Weinan Zhang.
\newblock Holos: A web-scale llm-based multi-agent system for the agentic web, 2026.
\newblock URL \url{https://www.holosai.io/static_files/Holos_Paper_20260118.pdf}.

\bibitem[Ritchie and Thompson(1974)]{ritchie1974unix}
Dennis~M. Ritchie and Ken Thompson.
\newblock The {UNIX} time-sharing system.
\newblock \emph{Communications of the ACM}, 17\penalty0 (7):\penalty0 365--375, 1974.

\bibitem[Jimenez et~al.(2024)Jimenez, Yang, Wettig, Yao, Pei, Press, and Narasimhan]{jimenez2024swebench}
Carlos~E. Jimenez, John Yang, Alexander Wettig, Shunyu Yao, Kexin Pei, Ofir Press, and Karthik Narasimhan.
\newblock {SWE-bench}: Can language models resolve real-world {GitHub} issues?
\newblock In \emph{International Conference on Learning Representations (ICLR)}, 2024.

\bibitem[Yang et~al.(2024)Yang, Jimenez, Wettig, Lieret, Yao, Narasimhan, and Press]{yang2024sweagent}
John Yang, Carlos~E. Jimenez, Alexander Wettig, Kilian Lieret, Shunyu Yao, Karthik Narasimhan, and Ofir Press.
\newblock {SWE-agent}: Agent-computer interfaces enable automated software engineering.
\newblock In \emph{Advances in Neural Information Processing Systems (NeurIPS)}, 2024.

\bibitem[{Foundation for Intelligent Physical Agents}(2002)]{fipa2002acl}
{Foundation for Intelligent Physical Agents}.
\newblock {FIPA} {ACL} message structure specification.
\newblock FIPA Standard SC00061G, 2002.
\newblock URL \url{http://www.fipa.org/specs/fipa00061/SC00061G.html}.

\bibitem[Smith(1980)]{smith1980contract}
Reid~G. Smith.
\newblock The contract net protocol: High-level communication and control in a distributed problem solver.
\newblock \emph{IEEE Transactions on Computers}, C-29\penalty0 (12):\penalty0 1104--1113, 1980.

\bibitem[Yang et~al.(2025{\natexlab{b}})Yang, Chai, Song, Qi, Wen, Li, Liao, Hu, Lin, Chang, Liu, Wen, Yu, and Zhang]{yang2025survey}
Yingxuan Yang, Huacan Chai, Yuanyi Song, Siyuan Qi, Muning Wen, Ning Li, Junwei Liao, Haoyi Hu, Jianghao Lin, Gaowei Chang, Weiwen Liu, Ying Wen, Yong Yu, and Weinan Zhang.
\newblock A survey of {AI} agent protocols.
\newblock \emph{arXiv preprint arXiv:2504.16736}, 2025{\natexlab{b}}.

\bibitem[Kong et~al.(2025)Kong, Lin, Xu, et~al.]{kong2025survey}
Dezhang Kong, Shi Lin, Zhenhua Xu, et~al.
\newblock A survey of {LLM}-driven {AI} agent communication: Protocols, security risks, and defense countermeasures.
\newblock \emph{arXiv preprint arXiv:2506.19676}, 2025.

\bibitem[Chang et~al.(2025)Chang, Lin, Yuan, Chen, et~al.]{chang2025anp}
Gaowei Chang, Eidan Lin, Chengxuan Yuan, Yifei Chen, et~al.
\newblock Agent network protocol technical white paper.
\newblock \emph{arXiv preprint arXiv:2508.00007}, 2025.

\bibitem[Wu et~al.(2023)Wu, Bansal, Zhang, Wu, Li, Zhu, Jiang, Zhang, Zhang, Liu, et~al.]{wu2024autogen}
Qingyun Wu, Gagan Bansal, Jieyu Zhang, Yiran Wu, Beibin Li, Erkang Zhu, Li~Jiang, Xiaoyun Zhang, Shaokun Zhang, Jiale Liu, et~al.
\newblock {AutoGen}: Enabling next-gen {LLM} applications via multi-agent conversation.
\newblock \emph{arXiv preprint arXiv:2308.08155}, 2023.

\bibitem[Li et~al.(2023)Li, Hammoud, Itani, Khizbullin, and Ghanem]{li2023camel}
Guohao Li, Hasan Abed Al~Kader Hammoud, Hani Itani, Dmitrii Khizbullin, and Bernard Ghanem.
\newblock {CAMEL}: Communicative agents for {``Mind''} exploration of large language model society.
\newblock In \emph{Advances in Neural Information Processing Systems (NeurIPS)}, volume~36, 2023.

\bibitem[Hong et~al.(2024)Hong, Zhuge, Chen, Zheng, Cheng, Zhang, Wang, Wang, Yau, Lin, et~al.]{hong2024metagpt}
Sirui Hong, Mingchen Zhuge, Jiaqi Chen, Xiawu Zheng, Yuheng Cheng, Ceyao Zhang, Jinlin Wang, Zili Wang, Steven Ka~Shing Yau, Zijuan Lin, et~al.
\newblock {MetaGPT}: Meta programming for a multi-agent collaborative framework.
\newblock In \emph{International Conference on Learning Representations (ICLR)}, 2024.

\bibitem[Qian et~al.(2024)Qian, Liu, Liu, Chen, Dang, Li, Yang, Chen, Su, Cong, et~al.]{qian2024chatdev}
Chen Qian, Wei Liu, Hongzhang Liu, Nuo Chen, Yufan Dang, Jiahao Li, Cheng Yang, Weize Chen, Yusheng Su, Xin Cong, et~al.
\newblock Chatdev: Communicative agents for software development.
\newblock In \emph{Proceedings of the 62nd Annual Meeting of the Association for Computational Linguistics (ACL)}, 2024.

\bibitem[Gao et~al.(2025)Gao, Li, Xie, Kuang, Yao, Qian, Ma, et~al.]{gao2025agentscope}
Dawei Gao, Zitao Li, Yuexiang Xie, Weirui Kuang, Liuyi Yao, Bingchen Qian, Zhijian Ma, et~al.
\newblock {AgentScope} 1.0: A developer-centric framework for building agentic applications.
\newblock \emph{arXiv preprint arXiv:2508.16279}, 2025.

\bibitem[Wang et~al.(2025{\natexlab{a}})Wang, Chen, Song, Zhang, Ai, Yang, and Shi]{wang2025symphony}
Ji~Wang, Kashing Chen, Xinyuan Song, Ke~Zhang, Lynn Ai, Eric Yang, and Bill Shi.
\newblock Symphony: A decentralized multi-agent framework for scalable collective intelligence.
\newblock \emph{arXiv preprint arXiv:2508.20019}, 2025{\natexlab{a}}.

\bibitem[Yu(2026)]{yu2026adaptorch}
Geunbin Yu.
\newblock {AdaptOrch}: Task-adaptive multi-agent orchestration in the era of {LLM} performance convergence.
\newblock \emph{arXiv preprint arXiv:2602.16873}, 2026.

\bibitem[Yu et~al.(2025)Yu, Yu, Wang, Wang, Yang, Li, Li, and Yang]{yu2025infiagent}
Chenglin Yu, Yang Yu, Songmiao Wang, Yucheng Wang, Yifan Yang, Jinjia Li, Ming Li, and Hongxia Yang.
\newblock {InfiAgent}: Self-evolving pyramid agent framework for infinite scenarios.
\newblock \emph{arXiv preprint arXiv:2509.22502}, 2025.

\bibitem[Huang et~al.(2025)Huang, Cheng, and Tan]{huang2025evogit}
Beichen Huang, Ran Cheng, and Kay~Chen Tan.
\newblock {EvoGit}: Decentralized code evolution via git-based multi-agent collaboration.
\newblock \emph{arXiv preprint arXiv:2506.02049}, 2025.

\bibitem[Li et~al.(2025)Li, Ping, Chen, Qi, Wang, Luo, and Zhang]{li2025agentgit}
Yang Li, Siqi Ping, Xiyu Chen, Xiaojian Qi, Zigan Wang, Ye~Luo, and Xiaowei Zhang.
\newblock {AgentGit}: A version control framework for reliable and scalable {LLM}-powered multi-agent systems.
\newblock \emph{arXiv preprint arXiv:2511.00628}, 2025.

\bibitem[Benkovich and Valkov(2026)]{benkovich2026agyn}
Nikita Benkovich and Vitalii Valkov.
\newblock Agyn: A multi-agent system for team-based autonomous software engineering.
\newblock \emph{arXiv preprint arXiv:2602.01465}, 2026.

\bibitem[Pugachev(2025)]{pugachev2025codecrdt}
Sergey Pugachev.
\newblock {CodeCRDT}: Observation-driven coordination for multi-agent {LLM} code generation.
\newblock \emph{arXiv preprint arXiv:2510.18893}, 2025.

\bibitem[Mao et~al.(2025)Mao, Keung, Zhang, et~al.]{mao2025semap}
Zhenyu Mao, Jacky Keung, Fengji Zhang, et~al.
\newblock {SEMAP}: Software engineering multi-agent protocol.
\newblock \emph{arXiv preprint arXiv:2510.12120}, 2025.

\bibitem[Pike et~al.(1995)Pike, Presotto, Dorward, Flandrena, Thompson, Trickey, and Winterbottom]{pike1995plan9}
Rob Pike, Dave Presotto, Sean Dorward, Bob Flandrena, Ken Thompson, Howard Trickey, and Phil Winterbottom.
\newblock Plan 9 from {Bell Labs}.
\newblock \emph{Computing Systems}, 8\penalty0 (3):\penalty0 221--254, 1995.

\bibitem[Sandberg et~al.(1985)Sandberg, Goldberg, Kleiman, Walsh, and Lyon]{sandberg1985nfs}
Russel Sandberg, David Goldberg, Steve Kleiman, Dan Walsh, and Bob Lyon.
\newblock Design and implementation of the {Sun Network Filesystem}.
\newblock In \emph{Proceedings of the USENIX 1985 Summer Conference}, pages 119--130, 1985.

\bibitem[Vangoor et~al.(2017)Vangoor, Tarasov, and Zadok]{vangoor2017fuse}
Bharath Kumar~Reddy Vangoor, Vasily Tarasov, and Erez Zadok.
\newblock To {FUSE} or not to {FUSE}: Performance of user-space file systems.
\newblock In \emph{15th USENIX Conference on File and Storage Technologies (FAST '17)}, pages 59--72. USENIX Association, 2017.

\bibitem[Wang et~al.(2025{\natexlab{b}})Wang, Li, Song, et~al.]{wang2025openhands}
Xingyao Wang, Boxuan Li, Yufan Song, et~al.
\newblock {OpenHands}: An open platform for {AI} software developers as generalist agents.
\newblock In \emph{International Conference on Learning Representations (ICLR)}, 2025{\natexlab{b}}.

\bibitem[Chen et~al.(2025)Chen, Peng, Yang, et~al.]{chen2025envx}
Linyao Chen, Zimian Peng, Yingxuan Yang, et~al.
\newblock {EnvX}: Agentize everything with agentic {AI}.
\newblock \emph{arXiv preprint arXiv:2509.08088}, 2025.

\bibitem[Hassan et~al.(2025)Hassan, Li, Lin, et~al.]{hassan2025sase}
Ahmed~E. Hassan, Hao Li, Dayi Lin, et~al.
\newblock Structured agentic software engineering.
\newblock \emph{arXiv preprint arXiv:2509.06216}, 2025.

\bibitem[{Anthropic}(2025)]{anthropic2025agentskills}
{Anthropic}.
\newblock Agent skills.
\newblock \url{https://platform.claude.com/docs/en/agents-and-tools/agent-skills/overview}, 2025.
\newblock Accessed: 2026-02-14.

\bibitem[{WHATWG}(2026)]{whatwg2026sse}
{WHATWG}.
\newblock Html living standard: Server-sent events.
\newblock \url{https://html.spec.whatwg.org/multipage/server-sent-events.html}, 2026.
\newblock Accessed: 2026-02-14.

\bibitem[{Cline Contributors}(2024)]{cline2024}
{Cline Contributors}.
\newblock Cline: An autonomous coding agent for vs code.
\newblock \url{https://github.com/cline/cline}, 2024.
\newblock Accessed: 2026-02-14.

\bibitem[Wu et~al.(2025)Wu, Roesner, Kohno, Zhang, and Iqbal]{wu2025isolategpt}
Yuhao Wu, Franziska Roesner, Tadayoshi Kohno, Ning Zhang, and Umar Iqbal.
\newblock {IsolateGPT}: An execution isolation architecture for {LLM}-based agentic systems.
\newblock In \emph{Network and Distributed System Security Symposium (NDSS)}, 2025.

\bibitem[Zhang et~al.(2026)Zhang, Liu, Liu, Zhao, Wang, Xu, Niyato, Kang, Li, Mao, et~al.]{zhang2026toward}
Ruichen Zhang, Guangyuan Liu, Yinqiu Liu, Changyuan Zhao, Jiacheng Wang, Yunting Xu, Dusit Niyato, Jiawen Kang, Yonghui Li, Shiwen Mao, et~al.
\newblock Toward edge general intelligence with agentic ai and agentification: Concepts, technologies, and future directions.
\newblock \emph{IEEE Communications Surveys \& Tutorials}, 28:\penalty0 4285--4318, 2026.

\bibitem[Yang and Zhu(2026)]{yang2026internetagentic}
Ya-Ting Yang and Quanyan Zhu.
\newblock Internet of agentic {AI}: Incentive-compatible distributed teaming and workflow.
\newblock \emph{arXiv preprint arXiv:2602.03145}, 2026.

\bibitem[Wooldridge et~al.(2025)Wooldridge, Bagoly, Ward, La~Malfa, and Paludo~Licks]{wooldridge2025fetchai}
Michael~J. Wooldridge, Attila Bagoly, Jonathan~J. Ward, Emanuele La~Malfa, and Gabriel Paludo~Licks.
\newblock Fetch.ai: An architecture for modern multi-agent systems.
\newblock \emph{arXiv preprint arXiv:2510.18699}, 2025.

\bibitem[Giusti et~al.(2025)Giusti, Werner, Taiello, Costa, Tosun, Protani, Molina, de~Almeida, Cacace, Santos, et~al.]{giusti2025federation}
Lorenzo Giusti, Ole~Anton Werner, Riccardo Taiello, Matilde~Carvalho Costa, Emre Tosun, Andrea Protani, Marc Molina, Rodrigo~Lopes de~Almeida, Paolo Cacace, Diogo~Reis Santos, et~al.
\newblock Federation of agents: A semantics-aware communication fabric for large-scale agentic {AI}.
\newblock \emph{arXiv preprint arXiv:2509.20175}, 2025.

\end{thebibliography}
